\documentclass[
showpacs,
floatfix,
aps,
prl,
twocolumn,
superscriptaddress,
amssymb,
]{revtex4-1}

\usepackage{graphicx}
\usepackage{bm}
\usepackage{color}
\usepackage[normalem]{ulem}

\usepackage{amsmath}
\usepackage{cleveref}

\usepackage{textcomp}

\newcommand{\be}{\begin{equation}}
\newcommand{\ee}{\end{equation}}
\newcommand{\bea}{\begin{eqnarray}}
\newcommand{\eea}{\end{eqnarray}}

\def\wse2{WSe$_2$}
\def\mose2{MoSe$_2$}
\def\moire{moir\'e }
\def\didv{$dI/dV$}

\begin{document}

\title{Deep \moire potentials in twisted transition metal dichalcogenide bilayers}

\author{Sara Shabani}
\affiliation{Department of Physics, Columbia University, New York, NY, USA}
\author{Dorri Halbertal}
\affiliation{Department of Physics, Columbia University, New York, NY, USA}
\author{Wenjing Wu}
\affiliation{Department of Chemistry, Columbia University, New York, NY, USA}
\author{Mingxing Chen}
\affiliation{School of Physics and Electronics, Hunan Normal University,
Key Laboratory for Matter Microstructure and Function of Hunan Province,
Key Laboratory of Low-Dimensional Quantum Structures and Quantum Control of Ministry of Education,
Changsha, Hunan, China}
\author{Song Liu}
\affiliation{Department of Mechanical Engineering, Columbia University, New York, NY, USA}
\author{James Hone}
\affiliation{Department of Mechanical Engineering, Columbia University, New York, NY, USA}
\author{Wang Yao}
\affiliation{Department of Physics and Center of Theoretical and Computational Physics, University of Hong Kong, Hong Kong, China}
\author{Dmitri N. Basov}
\affiliation{Department of Physics, Columbia University, New York, NY, USA}
\author{Xiaoyang Zhu}
\affiliation{Department of Chemistry, Columbia University, New York, NY, USA}
\author{Abhay N. Pasupathy}
\affiliation{Department of Physics, Columbia University, New York, NY, USA}

\date{\today}

\maketitle

\textbf{
In twisted bilayers of semiconducting transition metal dichalcogenides (TMDs), a combination of structural rippling and electronic coupling gives rise to periodic \moire potentials that can confine charged and neutral excitations\cite{Xu:2019:signature,MacDonald:2018,Yao:2017:emitter,Xu:2019:evidence,Augusto:2020,XYZ:2020:excitonic,PhilipKim:2019}. Here, we report experimental measurements of the structure and spectroscopic properties of twisted bilayers of \wse2 and \mose2 in the H-stacking configuration using scanning tunneling microscopy (STM). Our experiments reveal that the \moire potential in these bilayers at small angles is unexpectedly large, reaching values of above 300 meV for the valence band and 150 meV for the conduction band - an order of magnitude larger than theoretical estimates based on interlayer coupling alone. We further demonstrate that the \moire potential is a non-monotonic function of \moire wavelength, reaching a maximum at around a 13nm \moire period. This non-monotonicity coincides with a drastic change in the structure of the \moire pattern from a continuous variation of stacking order at small \moire wavelengths to a one-dimensional soliton dominated structure at large \moire wavelengths. We show that the in-plane structure of the \moire pattern is captured well by a continuous mechanical relaxation model, and find that the \moire structure and internal strain rather than the interlayer coupling is the dominant factor in determining the \moire potential. Our results demonstrate the potential of using precision \moire structures to create deeply trapped carriers or excitations for quantum electronics and optoelectronics.}

Lattice vector mismatches between two layers of a van der Waals bilayer gives rise to a \moire pattern. The \moire pattern affects the electronic structure of the bilayer, and many emergent quantum phenomena have recently been observed in these systems  \cite{Wang:2020:wigner,Augusto:2020,Xu:2019:evidence,KinFaiMac:2019}. In a TMD semiconductor heterobilayer, the low-energy electronic structure can be reasonably approximated by the properties of a single layer on which a spatially dependent potential energy landscape is imposed (termed the \moire potential)\cite{MacDonald:2018:Hubbard,LiangFu2019,MacDonald:2018}. This \moire potential when periodic gives rise to subbands within the first valence or conduction bands, which are responsible for the emergent quantum properties observed. Spatially separated interlayer excitons can also be trapped within these subbands\cite{Li:2019:exciton,Xu:2019:evidence,Xu:2019:signature,PhilipKim:2019,KinFaiMac:2019,XYZ:2020:excitonic}. Theoretical estimates based only on interlayer coupling estimate the size of this \moire potential to be of the order of a 10 millielectronvolts (meV) at small \moire wavelengths ($<5$ nm)\cite{Yao:2017:potential,Falko:2019}, but experimental measurements of the \moire potential remains an important open problem in these materials.
 
 Scanning tunneling microscopy is one of the few experimental techniques that can provide direct measurements of the magnitude of the \moire potential, due to its high energy and spatial resolution. Its use requires clean surfaces and conducting samples, both of which are significant challenges for TMD semiconductor layers. A few pioneering STM experiments have been performed on CVD grown \cite{Shih:2016,Feenstra:2018:confined}, rotationally aligned bilayers and (more recently) exfoliated, twisted TMD bilayers\cite{LeRoy:2019,Crommie:2020,Jonker:2020,Falko:2020:reconstruction}. All of these previous measurements have been performed for \moire wavelengths near 5 nm at rotational angles close to zero degrees between the two layers. In this work, we study the heterobilayer of \mose2 on \wse2 at a range of \moire wavelengths from 5- $>$20 nm. We avoid problems associated with sample conduction by performing our STM measurements at room temperature with a few-layer graphite substrate, under which condition the samples are sufficiently conductive. 
 
\begin{figure*}
\centering
\includegraphics[width=1\linewidth]{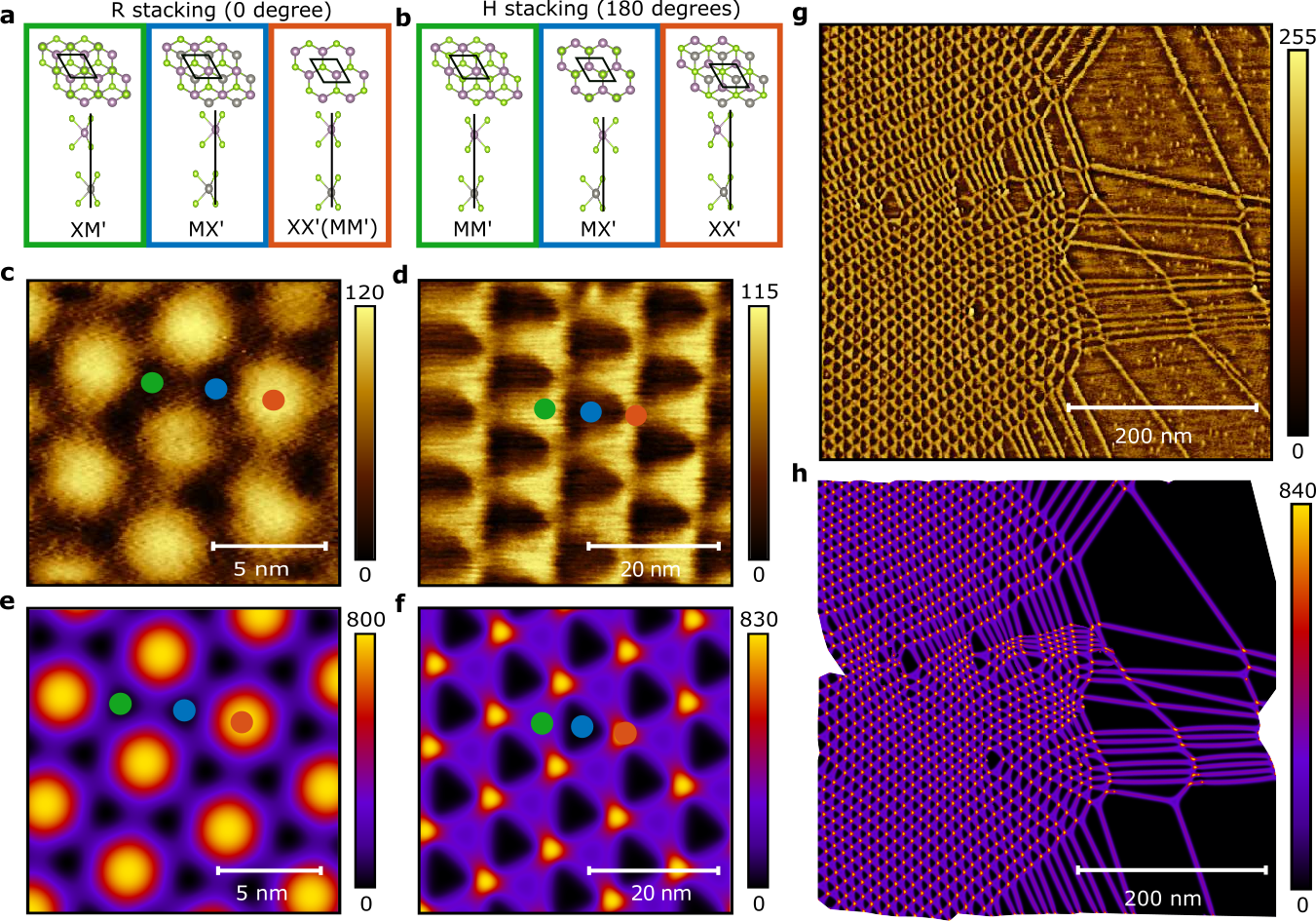}                                                                                                                                                
\vspace{-0.15 in}
\caption{\small{
{\bf{Structure of twisted heterobilayer \wse2/\mose2.}}
{\bf{a,b}} Illustration of R- and H- stacked heterobilayers and the high-symmetry stacking configurations present within each configuration.
{\bf{c,d}} STM topographic images (in $pm$) of R- and H-stacked twisted heterobilayers. (set point of -1.4 V, 100 pA and -1.7 V, 150 pA, respectively).  
{\bf{e,f}} Stacking energy density (in $meV/nm^2$) from mechanical relaxation calculations corresponding to the STM topographs in {\bf{c,d}}. The high symmetry stacking configurations illustrated in {\bf{a,b}} are marked with appropriately colored dots in {\bf{c-f}}. 
{\bf{g}} Large area STM topograph (in $pm$) of H-stacked \wse2/\mose2 at an average twist angle of $\sim$1.7$^{\circ}$. The topograph shows the presence of inhomogeneous heterostrain, one-dimensional solitons, point defects in the individual layers and edge dislocations of the \moire lattice. {\bf{h}} Calculated stacking energy density (in $meV/nm^2$) of the relaxed structure. The stacking registry was forced at selected points from the experimental picture. The image is composed of several separate calculations surrounding the observed dislocations from both sides (details in supplementary information).
}
}
\label{fig1}
\vspace{-0.15 in}
\end{figure*}

Due to the broken inversion symmetry in TMDs, there are two distinct aligned stacking configurations termed R (zero degree alignment between the two layers, also termed AA stacking in the literature) and H (180-degree alignment between the two layers, also termed AB stacking). When a twist angle is present between the layers, the atomic registry between the two layers varies periodically in space\cite{Yao:2017:emitter}. For nearly R-stacked twisted bilayers, three of the high-symmetry stacking orders that are present in the sample are shown in figure 1a, termed MM', MX', and M'X respectively. Here M and X refer to the metal and chalcogen atoms in the top layer, while M' and X' refer to those in the bottom layer; MM' refers to the stacking where the metal atoms of the top layer are in vertical registry with the metal atoms from the bottom layer. For nearly H-stacked bilayers, the corresponding high-symmetry stackings are XX', MX'and MM' as shown in figure 1b. 

\begin{figure*}
\centering
\includegraphics[width=1\linewidth]{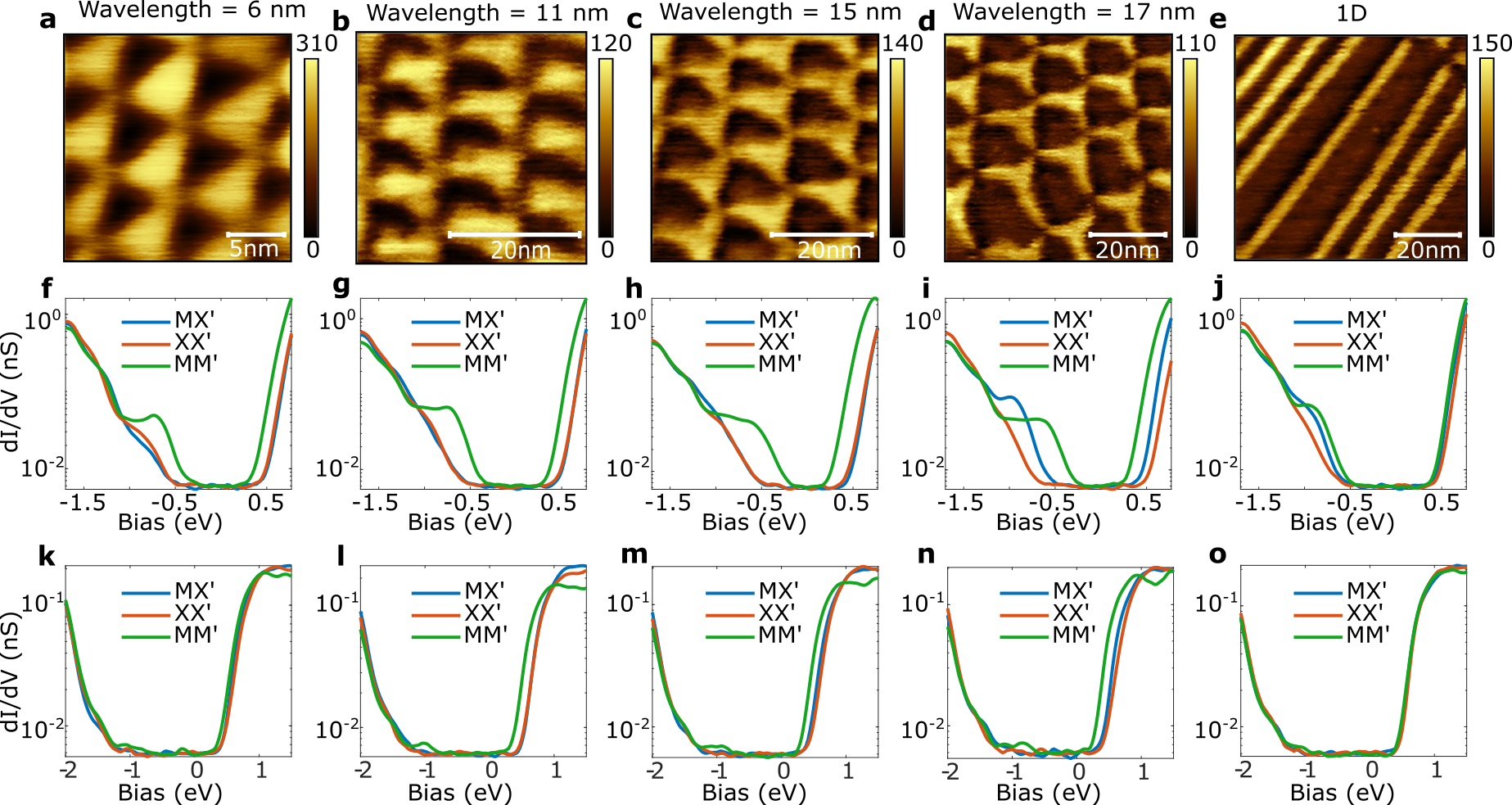}
\vspace{-0.15 in}
\caption{\small{
{\bf{Spectroscopic properties of \moire patterns of different wavelengths}}
{\bf{a-e}} STM topographic images (in $pm$) of \moire patterns of different wavelength (set points of -1.7 V and 100 pA). As the wavelength is increased, the area occupied by the MM' stacking configuration (brighter area) reduces, leading from a transition from a triangular lattice at small wavelength to a strain-soliton structure at large wavelength.
{\bf{f-j}} dI/dV measurements obtained at the high symmetry stacking configurations for each of {\bf{a-e}}. It is clearly seen that the MM' stacking configuration displays the smallest band gap, with both the conduction and valence edges shifted towards the Fermi level relative to the M-X' stacking configuration.
{\bf{k-o}} dI/dV measurements obtained over a larger energy range. We can see the band edge of the conduction band clearly, and at high negative energies the states from the \mose2 valence bands that dominate the tunneling. The valence bands nearest the Fermi level that are seen clearly in {\bf{f-j}} are much smaller in conductance on this scale, and are not seen clearly.
}}
\label{fig2}
\vspace{-0.15 in}
\end{figure*}
Shown in figure 1c and 1d are STM topographs of samples at twist angles of $\sim$3$^{\circ}$ (near R stacking) and $\sim$61.7$^{\circ}$ (near H stacking) respectively. We can see that the two stacking orientations present very different structures as visualized in STM. In order to understand this difference, we calculate the relaxed structure of twisted bilayers based upon a continuous mechanical relaxation model (details in methods)\cite{Dorri:2020,Jain:2019:reconstruction,Falko:2020:domain}. The results of these calculations are shown in figure 1e and f for the same angles as shown in figure 1c,d. For samples near R stacking, the MM' stacking is an energetic maximum, while both the MX' and M'X stackings are minima that are close to each other in energy (in our case, the chemical identity of M=Mo, M'=W, X=X'=Se).Consequently, the twisted bilayer shows topographic contrast with the MM' region (red dot, figure 1c) displaying a larger topographic height while the MX' (green dot) and M'X (blue dot) regions are minima in height. For samples near H stacking, the XX' stacking is most energetically unfavorable, while the MX' is the global minimum and the MM' is a local minimum. As a result, the XX' stacking (red dot, figure 1d) region shrinks in size while the MX' (blue dot, figure 1d) region expands and the MM' has an intermediate area (green dot, figure 1d). We also see quite clearly for this \moire wavelength ($\sim$11 nm) that lattice reconstructions result in fairly sharp triangular domain boundaries between the MX' and MM' regions. All of these together allow us to clearly identify the various stacking configurations in our STM topograph in figure 1d, as indicated in the figure. We see that samples that are near H stacking present a completely different structure than samples near R stacking, which has not been studied by STM previously. For the rest of this work, we focus exclusively on this case. 
  

\begin{figure*}
\centering
\includegraphics[width=1\linewidth]{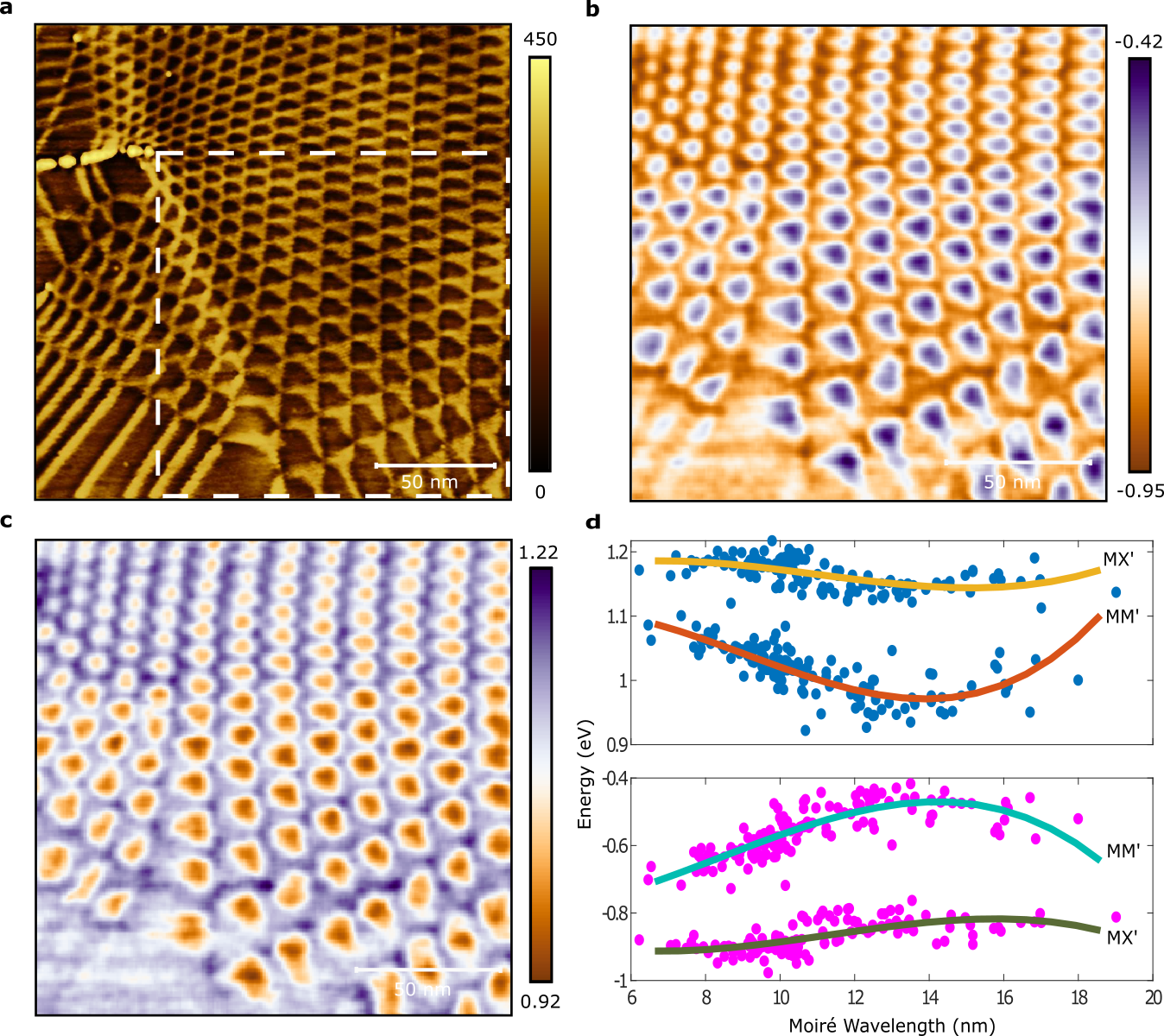}
\vspace{-0.15 in}
\caption{\small{
{\bf{Spectroscopic imaging of conduction and valence band edges}}
{\bf{a}} Topograph (in $pm$) of a non uniform  moire region (set points of -1.7V and 100 pA)
{\bf{b,c}} Valence and Conduction band edge maps (in $eV$) obtained in a uniform 256x256 grid within the dashed  box shown in a.
{\bf{d}} Extracted positions of the band edges in the MM' and MX' configurations from the maps in {\bf{b,c}}. These are plotted as a function of \moire wavelength calculated from the area of each \moire unit cell (see supplement for details). The solid lines are fifth order fits to the data.
}}
\label{fig3}
\vspace{-0.15 in}
\end{figure*}
 Having understood the details of the \moire pattern at small length scales, we proceed to perform STM measurements over large areas of nearly H-stacked samples. One such topograph is shown in figure 1g, over an area of 500x500 $nm^2$. This topograph shows many interesting features, including a spatially varying \moire period, large ($>$ 100 nm) regions of uniform MX' stacking, and one-dimensional solitons. Interesting electronic and optical properties have been reported in these 1D solitons\cite{Drew:2020,XYZ:2019:1d,Jain:2018:Soliton}. Point defects in the individual layers are seen as white dots of atomic dimensions, and edge dislocations in the \moire lattice are also observed, presumably due to impurities between the two layers of the bilayer. These features arise from the presence of impurities and non-uniform strain over this area. All of this disparate behavior can be captured with the continuum mechanical relaxation calculation shown in figure 1h. The only inputs (beyond those of the periodic case shown in figure 1f) that go into this calculation are the locations and stacking registry of selected XX' stacking points. Given this information, a detailed spatial account of the relaxed structure is resolved. Due to the existing dislocations, the process was repeated surrounding dislocations from different sides to generate the integrated map of figure 1h (see methods and supplementary information). We find that this process captures the entire complex structure of the \moire pattern, and can be used to quantitatively estimate the local strain fields producing inhomogeneities in the large scale structure in figure 1g.


\begin{figure*}
\centering
\includegraphics[width=1\linewidth]{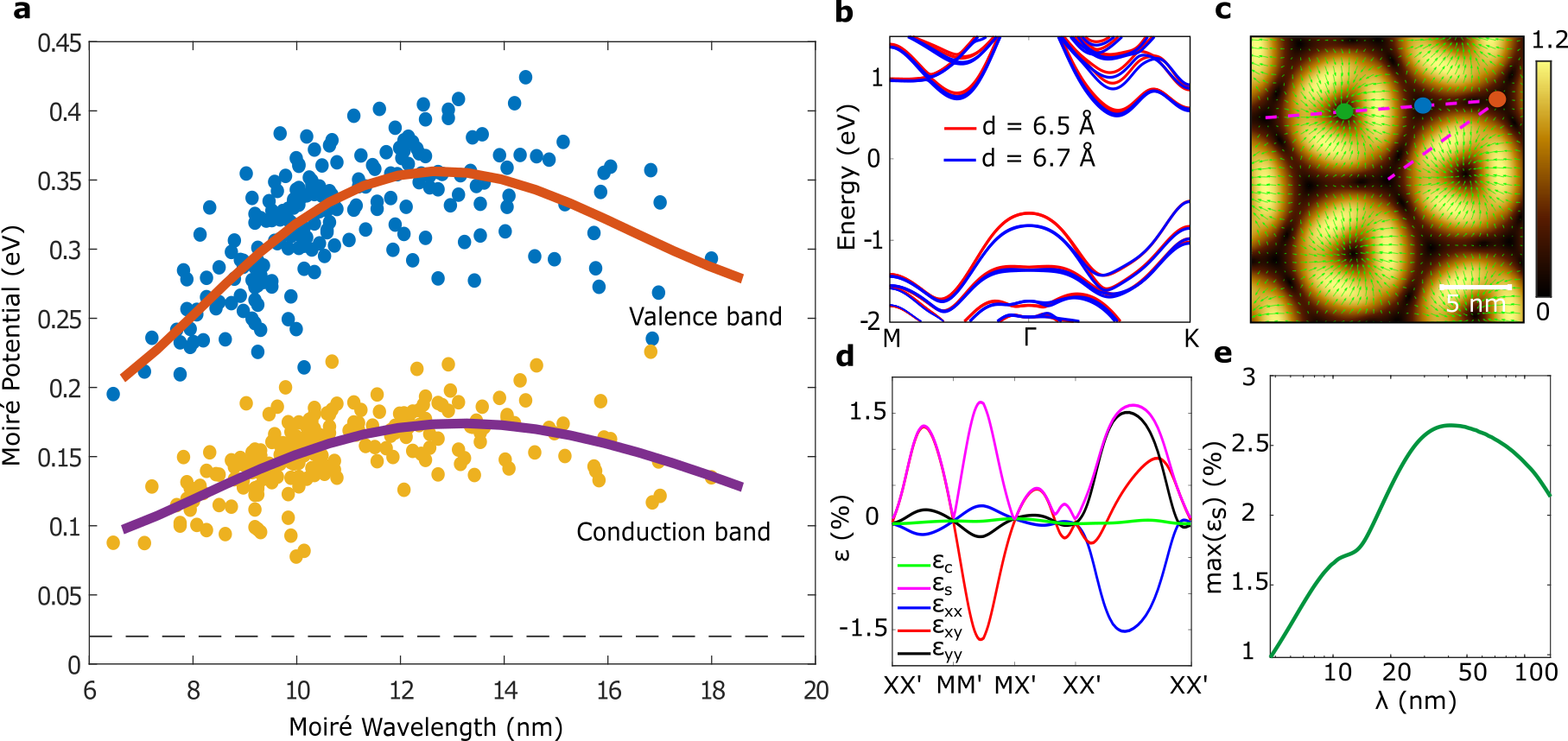}
\vspace{-0.15 in}
\caption{\small{
{\bf{Quantifying the \moire potential}}
{\bf{a}} \moire potential as a function of \moire wavelength extracted from spectroscopic imaging experiments. Solid lines are fifth order fits to the data. The dashed line is the theoretically calculated difference in valence band edges between uniformly stacked MM' and MX' bilayers. It is seen that the experimentally measured \moire potentials are much larger than this theoretical expectation. 
{\bf{b}} DFT bandstructure of uniform MM' stacked bilayers with the equilibrium layer separation (blue) and for a bilayer that is compressed by 0.2 $\AA$.   
{\bf{c}} Shear strain map ($\epsilon_s$ in \%) from mechanical relaxation calculations of H-stacked \mose2/\wse2 with a \moire period of 10 nm. The green arrows make a stream plot of the $\epsilon_s\cdot(\cos{\phi},\sin{\phi})$ field (see text for definitions of $\epsilon_c$, $\epsilon_s$ and $\phi$), providing additional information about the shear direction. Colored dots in {\bf{c}} correspond to stacking configuration marked in Fig. 1. {\bf{d}} Strain tensor components ($\epsilon_{xx}$, $\epsilon_{xy}$=$\epsilon_{yx}$, $\epsilon_{yy}$) and $\epsilon_c$, $\epsilon_s$ along the path marked by the dashed lines in {\bf{c}}. {\bf{e}} Maximal shear strain as a function of \moire wavelength, showing a non-monotonic behavior.
}}
\label{fig4}
\vspace{-0.15 in}
\end{figure*}
We now proceed to examine the structure of the \moire pattern at various length scales. Shown in figure 2a-e are a sequence of topographs obtained in regions with increasing \moire length scales. At the smallest of these length scales ( $\sim$ 6 nm, figure 2a), the \moire pattern features nearly equal regions of MX' and MM' stacking. As the size of the \moire wavelength increases (figure 2b-d), the area of the MM' stacked region decreases at the expense of the area of the MX' stacking region. For \moire wavelengths that are larger than 20 nm, the MM' region shrinks to a shear soliton of width approximately $\sim$ 4 nm. Above this wavelength, \moire patterns resemble honeycomb lattices formed by the shear solitons rather than the triangular lattices seen at small wavelengths (see supplementary information for a typical image). The large \moire wavelengths are extremely susceptible to small amounts of strain, which can distort the honeycomb structure severely. An example is shown in figure 2e, where the individual honeycomb cells have been distorted to form quasi-rectangular strips of MX' stacking that are separated by soliton domain walls.

We now consider the spectroscopic properties of samples exhibiting different \moire wavelengths. Shown in figure 2f-j are measurements of the differential conductance (\didv) obtained at the three high symmetry locations of the \moire lattice, viz. XX', MM' and MX' for the \moire patterns shown in figure 2a-e respectively. Clear and systematic differences are seen in the spectroscopic properties of the different sites within the \moire unit cell. It is seen that the edge of both the valence band and the conduction band are closest to the Fermi level for the MM' site for all of the \moire wavelengths. The difference in valence band edges between the MM' and MX' regions reaches a maximum at a \moire wavelength around 13 nm(figure 2c), and decreases for both smaller and larger \moire wavelengths. Similar behavior is observed for the conduction band edges. The wavelength at which the \moire potential is largest corresponds structurally to the length scale where the MM' region transitions from a triangular region to a soliton. The valence band edge observed in figure 2f-j is derived from the states with primarily \wse2 character, while the conduction band edge states are derived from states with primarily \mose2 character\cite{Yao:2017:emitter}. The states with \wse2 character have a small tunnel matrix element due to the larger physical distance from the STM tip. The conduction band states, therefore, have a much higher intensity than the valence band states shown in figure 2f-j. Spectra taken over a wider bias range, shown in figures 2k-o show clearly the conduction band edges as well as deeper valence band states that are derived from the \mose2 layer. We use these spectra to define the edges of the conduction and valence bands (see supplementary information for more details of the procedure).

The spectroscopic differences between the various regions of the \moire unit cell described above are easiest to understand by considering a single \moire unit cell of a single layer. Within this unit cell, a \moire potential energy exists that shifts the location of the band edge. Thus, the \moire unit cell can simply be considered to be a problem of a triangular quantum well with finite depth. Within a single unit cell, this gives rise to a number of confined quantum dot states \cite{Feenstra:2018:confined,Jain:2019:reconstruction}. At low energy, the states are localized inside the well while at energies above the well depth, the states are found outside the well. At room temperature, we average over closely spaced states and instead see a band edge both inside and outside the quantum well. The difference in the band edge positions inside and outside the well is then simply equal to the well depth, ie, the magnitude of the \moire potential. Our spectroscopic results indicate that the MM' region of the \moire unit cell is the region with a potential minimum for both the valence and conduction band, ie, the low energy physics of this system is dominated by electrons or holes trapped within these regions. A cursory inspection of figure 2f-j also reveals that this trapping potential is large - around 300 meV at its largest in the valence band, and 100 meV for the conduction band. This consideration becomes especially interesting for large \moire wavelengths, where the MM' regions shrink to soliton lines. Our results indicate that carriers are confined to these one-dimensional lines at low energies in these structures. This quantum well picture that is based on a single \moire unit cell is only slightly modified by the periodic boundary conditions imposed by the \moire pattern - the coupling between neighboring wells broadens each of the eigenstates within the well to a "flat band" with width determined by inter-well coupling\cite{MacDonald:2018:Hubbard}.

In order to understand the systematic evolution of the \moire potential as a function of \moire wavelength, we can utilize strain-induced inhomogeneity in the sample to our advantage. Shown in figure 3a is a region of the sample where the \moire wavelength interpolates continuously between a minimum of 5 nm to ($>$ 20 nm). We proceed to take \didv spectroscopic measurements at every point on a 256x256 pixel grid in the area bounded by the dashed box in figure 3a. For each spectrum, we determine the edge of the valence band and the edge of the conduction band and plot their values in figures 3b and c respectively. We use this information, together with the areas of the individual \moire unit cells to plot the conduction and valence band edges as a function of \moire wavelength in figure 4d (for details of the analysis, see supplementary information). The scatter in this figure primarily arises from the fact that different \moire triangles in figure 3a have differing shapes and thus display slightly different properties when represented by a single length scale. The trends shown in figure 3d confirm the selected point spectra shown in figure 2 - the non-monotonicity of the band edges as a function of \moire wavelength from figure 2 is clearly confirmed in figure 3d in a much more extensive data set.

Having obtained separately the conduction and valence band edges for the MM' and MX' sites in figure 3, we can utilize this information to extract the \moire potential as a function of wavelength, which is simply the energy difference in band positions between a MM' region and its MX' neighbors(see supplementary information for details of the analysis). This is plotted in figure 4a for both the conduction and valence band. We can see clearly that the \moire potential is large and non-monotonic for both the conduction and valence band edges. In previous theoretical considerations, the hybridization between the two layers of the heterostructure has been considered to play a dominant role in the \moire potential\cite{Wang:2013:mos2}. For \wse2/\mose2, the hybridization differences between bilayers with uniform MM' and MX' stacking order are small (of order 10 meV), and thus there has been a theoretical expectation that the \moire potential is also similarly small in magnitude. Our results indicate that the true \moire potential is far larger than estimations based on uniform stacking order. Some insight into this difference can be gained from previous STM  experiments on \moire patterns in TMD bilayers \cite{Shih:2016,Feenstra:2018:confined,Crommie:2020,Cobden:2017}. In all of these works, the observed \moire potential is significantly larger than the expectation based on stacking order alone. This is the case even though these various experiments are on different materials from ours and also are in the R stacking configuration. The common feature of all of the works (including ours) is that real \moire patterns feature structural distortions in both the lateral and vertical dimensions - these distortions must, therefore, be dominating the \moire potential. 

Distortions within the \moire unit cell give rise to significant lateral and vertical strain. Due to the large unit cell sizes, we consider in our work, accurate ab-initio calculations of the electronic properties of the \moire are not feasible. We expect significant vertical strain to be present in the structure. For uniformly stacked bilayers, the MM' stacking configuration displays a significantly larger c-axis lattice constant in comparison to MX'(6.7$\AA$ instead of 6.2$\AA$). We thus expect that within the \moire structure, the MM' regions are being compressed down by the MX' regions, while the MX' regions are under tensile strain along the c axis. These strains can give rise to significant changes in electronic structure. Shown in figure 4b is a DFT calculation of mechanical relaxation model of a uniform MM' stacked region at an interlayer distance of 6.7 A (equilibrium) and 6.5 A (compressed). This compression gives rise to about a 0.2 eV shift in the valence band position at the $\Gamma$ point in a direction that is consistent with the experimental finding. We also expect significant lateral strains to be present within the unit cell, which we estimate using our mechanical relaxation model within the \moire unit cell. We write the strain tensor as ${\bf{\epsilon}}=\epsilon_C {\bf{I}} + \epsilon_S(\cos{(2\phi)}{\bf{\sigma_Z}}+\sin{(2\phi)}{\bf{\sigma_x}})$ where $\epsilon_C$ is an isotropic compression and $\epsilon_S$ is a volume-preserving shear, and $\bf{I}$ is the identity matrix and $\bf{\sigma_Z},\bf{\sigma_x}$ are the Pauli matrices in standard notation. The magnitude $\epsilon_S$ is plotted within the \moire unit cell in figure 4c for a \moire wavelength of $\sim$ 10 nm ($\epsilon_C$ is small at this wavelength). The total variation in strain across the unit cell is seen to be $>$ 3\% as can be seen in the line cut of strain tensor elements shown in figure 4d. For comparison, a uniform tensile strain of a percent changes the band gap of TMDs by $\approx$ 0.2 eV \cite{Drew:2020}. It is thus no surprise that these large values of strain in the \moire unit cell dominate the electronic properties. Fig. 4e shows the maximal value of $\epsilon_s$ as a function of the \moire wavelength from mechanical relaxation calculations. It is interesting to note that the shear strain also has a non-monotonic behavior as a function of \moire wavelength. This similarity in behavior to the experimentally observed \moire potential further supports the hypothesis that relaxation induced strain is the source of the observed enhanced \moire potential.
Our results show that the \moire potential in TMD heterobilayers is substantially larger than previous expectations, and can reach values of several hundred millivolts. Such large trapping potentials can be extremely useful in confining charge carriers as well as excitons and enhancing interactions between them. At the same time, our results show that the largest \moire potentials are realized for a narrow range of angles, and engineering high-quality structures with uniform \moire lattices with these wavelengths remains an open problem.

\bibliography{biba.bib}

\section{Acknowledgement}
We thank Stephen Carr and Efthimios Kaxiras for providing the GSFE function for relaxation calculations. We thank Cory Dean, Liang Fu, Manish Jain, Indrajit Maiti, Qianhui Shi, and Yang Zhang for useful discussions. This work is supported by the Programmable Quantum Materials (Pro-QM) program at Columbia University, an Energy Frontier Research Center established by the Department of Energy (grant DE- SC0019443). STM experiments were supported by the Air Force Office of Scientific Research via grant FA9550-16-1-0601 (SS, ANP). Synthesis of \mose2 and \wse2 was supported by the National Science Foundation Materials Research Science and Engineering Centers program through Columbia in the Center for Precision Assembly of Superstratic and Superatomic Solids (DMR-1420634). DH was supported by a grant from the Simons Foundation (579913). DFT calculations were performed by the support of the National Natural Science Foundation of China (Grant Nos. 11774084 and U19A2090, MC). 

\section{Author Contributions}
SS performed STM experiments, assisted by SL. DH performed relaxation calculations. WW performed SHG experiments. MC performed DFT calculations. JH, WY, DB, XYZ, and ANP advised. Data analysis and manuscript preparation were performed by SS and ANP with input from all coauthors.

\section{Additional information}

\section{Methods}

\subsection{Device fabrication}

Monolayers of WSe2 and MoSe2 were obtained by mechanical exfoliation from self-flux grown
bulk crystals\cite{Drew:defect}. The relative orientation between two TMD monolayers was determined by second harmonic generation(SHG).
We used polypropylene
carbonate (PPC) to pick up an h-BN flake, few-layer graphite, and \wse2 and \mose2 monolayers, respectively, using a high-precision rotation stage. In the final stage, the sample was flipped on a Si/SiO2 substrate at elevated temperature 120$^{\circ}$ C.

\subsection{STM Measurements}

STM and STS data were acquired at room temperature in ultra high vacuum conditions. A lock-in amplifier with modulation of 25 meV and 917 Hz was used for \didv spectroscopy measurements. 
\subsection{Second-Harmonic-Generation Measurement}
SHG measurements were used to determine the crystal orientations of WSe2 and MoSe2 monolayers. Linearly polarized femtosecond laser light (Spectrum Physics Tsunami, 80MHz, 800nm, 80 fs) was focused onto a monolayer with a ×100 objective (Olympus LMPLFLN100X). The reflected SHG signal at 400nm was collected using the same objective and detected by a photomultiplier tube (Hamamatsu R4220P) and recorded with a photon counter (BK PRECISION 1823A 2.4GHz Universal Frequency Counter). CVD grown triangular shape monolayer MoS2 (2D Layer) were used to calibrate the SHG setup

\subsection{DFT Calculation}
We use a slab structure to model the WSe2/MoSe2 heterostructure. To avoid artificial interactions between the polar slabs, we place two oppositely oriented WSe2/MoSe2 units with the mirror symmetry in the slab. Each slab is separated from its periodic images by 15 Å vacuum regions. Our DFT calculations were performed using the Vienna ab initio Simulation Package\cite{Kresse:1996}. We use the projector augmented wave method to construct pseudopotentials\cite{Kresse:1999}. The plane-wave energy cutoff is 400 eV. The exchange correlation functional is approximated by the generalized gradient approximation as parametrized by Perdew, Burke, and Ernzerhof\cite{Perdew:1996}. The Brillouin zone is sampled by a 30 x 30 x 1 k-mesh.  Van der Waals dispersion forces between the two constituents were taken into account using the optB88-vdW functional within the vdW-DF method developed by Klimeš et al \cite{DFT:2011:Angelos}.

\subsection{Atomic Relaxation Simulation}
Modeling of the atomic relaxation of twisted \mose2/\wse2 was performed within a continuity model
following the method presented in \cite{carr:2018:relaxation}, but solved in real space. In this model, the total energy of the
system is taken as the sum of elastic energy and a stacking energy term. The total energy was
minimized in search for the inter-layer real space displacement field corresponding to the relaxed structure. The
stacking configuration at selected XX'-stacking points were imposed as boundary condition as to account
for a specific case under study. In the periodic case (Fig. 1e-f and Fig. 4c-d) four such boundary conditions were used
to impose a fixed external strain condition. In the case of a non-uniform strain map as in Fig. 1f, such
points were needed wherever the \moire superlattice deviated from a uniformly periodic structure.
Special care was needed to describe the 11 dislocations in the image. For each dislocation the structure
was relaxed using two registry maps for the XX'-stacking cites, as to describe both sides of the
dislocation. The stacking energy maps were later stitched to for Fig. 1h. More details about the real
space atomic relaxation simulations specific to fig. 1h are presented in the supplementary information.
The mechanical relaxation parameters for the \mose2/\wse2 heterostructures were calculated using DFT as implemented in the Vienna \textit{ab initio} simulation package (VASP) version 5.4.4 \cite{Kresse:1996}.
All geometries included a vertical ($c$-axis) of 25 \AA{} to ensure no interaction between periodic images.
The colinear spin-polarized electronic structure was calculated with a plane-wave cutoff of 500 eV, the VASP PAW PBE potentials (v54) \cite{Kresse:1999}, a broadening of 50 meV, and a self-consistency convergence criterion of $10^{-6}$ eV.
A periodic dipole correction in the $c$-axis to the total energy was included, and the van der Waals functional DFT-D3 (V3.0) was used \cite{Grimme:2010:consistent}.
The bulk modulus ($K$) and shear modulus ($G$) for each material were calculated by applying isotropic or uniaxial strain to a monolayer lattice, ranging from $-1.5\%$ to $1.5\%$ in units of $0.3\%$, and then performing a quadratic fit to the strain-dependent energies.
The generalized stacking fault energy function (GSFE) coefficients are extracted from a $6 \times 6$ sampling of the configuration between layers, with the vertical positions of the atoms relaxed at each configuration until all forces are less than 20 meV.
The Fourier components of the resulting energies are then extracted to create a convenient functional form for the GSFE used to describe the stacking energy term in the atomic relaxation calculations. 
The resulting GSFE coefficients and elastic coefficients used for the mechanical relaxation calculations were (following nomenclature of \cite{carr:2018:relaxation} and units of $\frac{meV}{u.c.}$): 
\mose2: $K$=40521, $G$=26464
\wse2: $K$=43113, $G$=30770
$c_0$=42.6, $c_1$=16.0, $c_2$=-2.7, $c_3$=-1.1, $c_4$=3.7, $c_5$=0.6
The unit-cell spacing are $\alpha$=0.3288 nm for \mose2 and $\alpha$=0.3282 nm for \wse2.
In all the mechanical relaxation calculations we assumed for simplicity one layer (\wse2) to be rigid, and allow all the relaxation to happen at the other (\mose2) layer. Relaxing this condition would not affect the overall picture significantly.
The periodic mechanical calculations of Fig. 1e, Fig. 1f and Fig. 4c-d assumed twist angle of 4$^{\circ}$, 1.5$^{\circ}$, and 1.88$^{\circ}$ and external strains of 0.13\%, 0.3\%, and 0\% respectively, all with a Poisson ratio of 0.23.

\end{document}


\title{Supplementary Information for \\ Deep \moire potentials in twisted transition metal dichalcogenide bilayers}

\author{Sara Shabani}
\affiliation{Department of Physics, Columbia University, New York, NY, USA}
\author{Dorri Halbertal}
\affiliation{Department of Physics, Columbia University, New York, NY, USA}
\author{Wenjing Wu}
\affiliation{Department of Chemistry, Columbia University, New York, NY, USA}
\author{Mingxing Chen}
\affiliation{School of Physics and Electronics, Hunan Normal University,
Key Laboratory for Matter Microstructure and Function of Hunan Province,
Key Laboratory of Low-Dimensional Quantum Structures and Quantum Control of Ministry of Education,
Changsha, Hunan, China}
\author{Song Liu}
\affiliation{Department of Mechanical Engineering, Columbia University, New York, NY, USA}
\author{James Hone}
\affiliation{Department of Mechanical Engineering, Columbia University, New York, NY, USA}
\author{Wang Yao}
\affiliation{Department of Physics and Center of Theoretical and Computational Physics, University of Hong Kong, Hong Kong, China}
\author{Dmitri N. Basov}
\affiliation{Department of Physics, Columbia University, New York, NY, USA}
\author{Xiaoyang Zhu}
\affiliation{Department of Chemistry, Columbia University, New York, NY, USA}
\author{Abhay N. Pasupathy}
\affiliation{Department of Physics, Columbia University, New York, NY, USA}

{
\let\clearpage\relax
\maketitle
\vspace{-10 ex}
}

\section{S1. Details of calculation of Figure 1h}

Figure 1h presents a mechanical relaxation calculation aiming to reproduce different non-uniform related strain features that were measured in figure 1g. Beyond the intrinsic system properties, namely the generalized stacking fault energy function (GSFE) lattice constants and elastic properties for this system (see methods section), the simulation uses as initial and boundary conditions the locations and stacking registry of selected XX' stacking points. Here we provide further details about this process. Figure S1 presents the initial and boundary conditions used to construct figure 1h. The white circles mark locations of dislocations appearing in the measurement of figure 1g. The dislocations are lattice defects, and are not directly supported by the model. However, locally the simulation can still account for the strain maps and stacking configurations. Therefore, the calculation was divided into several separate calculations, surrounding the dislocations from different orientations (region marked by dashed lines with different colors). In each region, the colored dots mark the positions of points where XX’ stacking configurations were forced for a given simulation. The false-color shows the stacking energy density of the initial configuration, used as a starting point for the simulations, which was generated as an interpolation between the forced stacking configurations. The solution of figure 1h is a result of stitching these 5 calculations.

\begin{figure}[hbp]
\centering
\includegraphics[height=9cm]{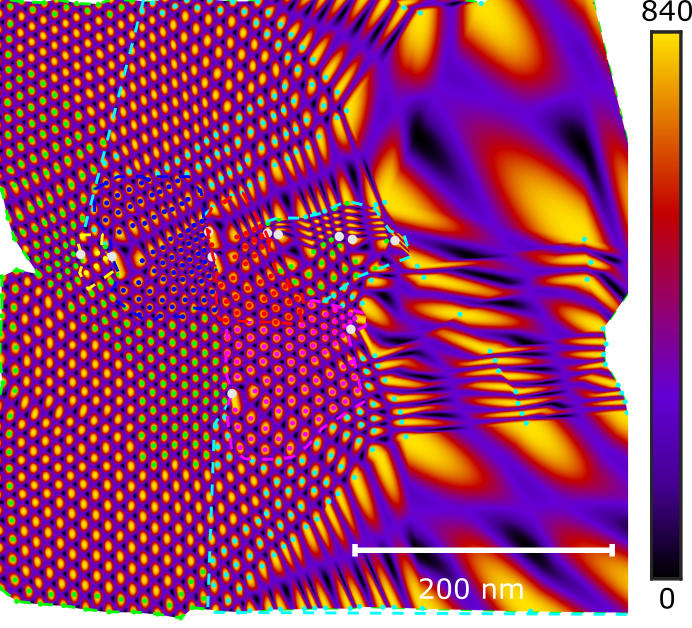}
\vspace{-0.15 in}
\caption{\small{ Initial and boundary conditions for the simulation presented in figure 1h. The white circles mark locations of dislocations appearing in the measurement of figure 1g. The simulation area was divided into 5 regions, marked by dashed lines with different colors. In each region, the colored dots mark the positions of points where XX’ stacking configurations were forced. The false-color shows the stacking energy density (in $meV/nm^2$) of the initial configuration, used as a starting point for the simulations. The solution of figure 1h is a result of stitching these 5 calculations.
}}
\captionsetup{labelformat=empty}
\vspace{-0.15 in}
\end{figure}

\section{S2. Transition from triangular to honeycomb structures in the \moire pattern}

At  \moire wavelengths of close to 10 nm near H-stacking, the \moire unit cell shows regions of MX and MM' stacking which are approximately equal in area, resulting in a triangular lattice \moire pattern as shown in figure 1d of the main text. As the wavelength increases beyond $\sim$ 15 nm, the MM' regions shrink in size to soliton lines, with nearly the entire \moire unit cell occupied by the MX' stacked regions. The \moire pattern then resembles a honeycomb lattice. The transition between the triangular and honeycomb regions is seen in figure S2. 

\begin{figure}[hbp]
\centering
\includegraphics[height=8cm]{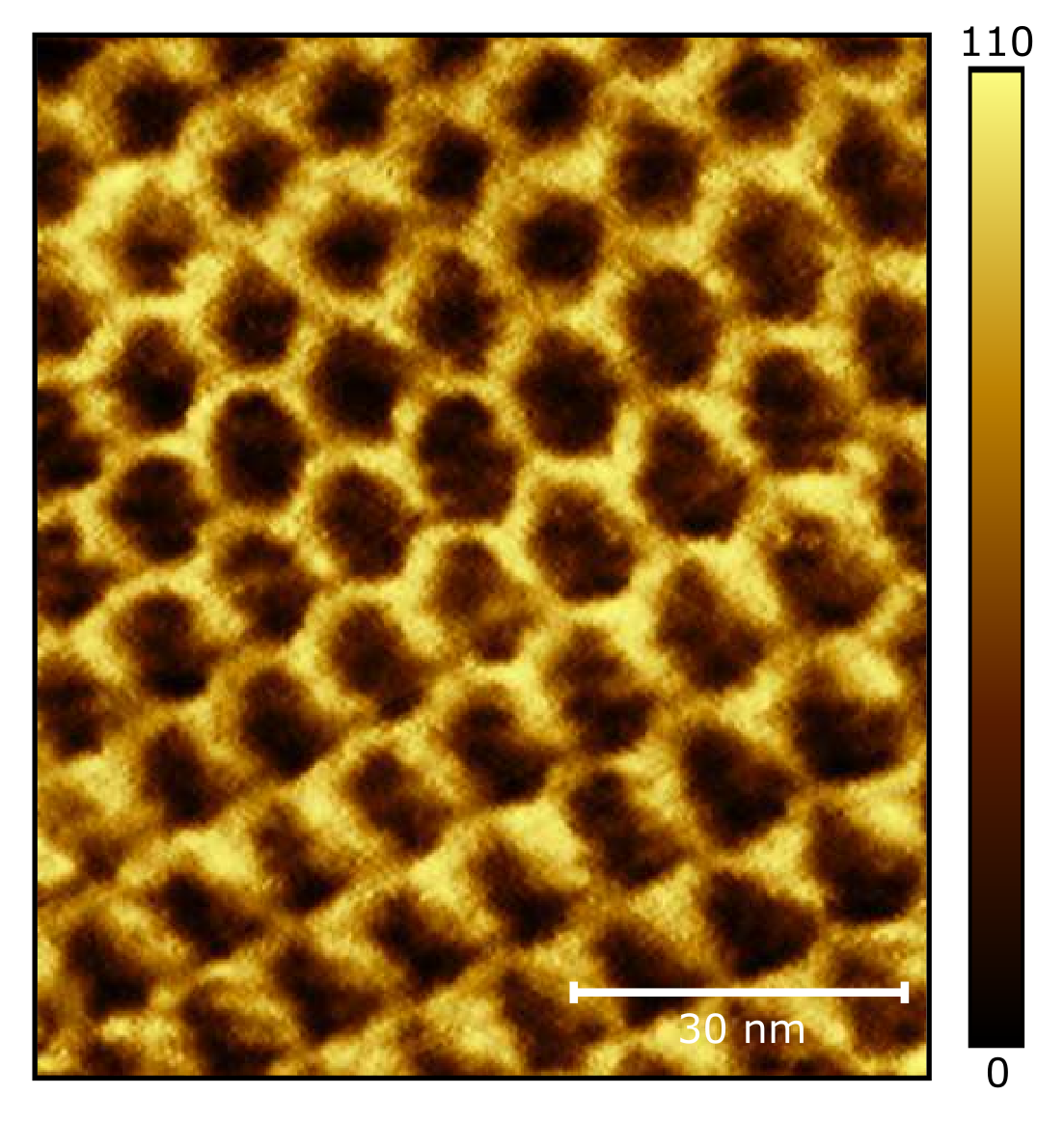}
\vspace{-0.15 in}
\caption{\small{ Honeycomb lattices formed by the shear strain instead of triangular lattice seen at small angle (in $pm$).
}}
\captionsetup{labelformat=empty}
\vspace{-0.15 in}
\end{figure}

\section{S3. Band edge analysis}

In two dimensions, the density of states $\rho_S(E)$ of an ideal semiconductor with a single valence and conduction band at T=0 features sharp band edges with a constant value of the DOS beyond the band edge. Our experiments are carried out at room temperature, which results in a broadening of the edges due to the Fermi distribution of electrons in the tip and sample at non-zero temperature. The expression for the differential conductance of the tip-sample junction is given by

\begin{figure}[hbp]
\centering
\includegraphics[width=\textwidth]{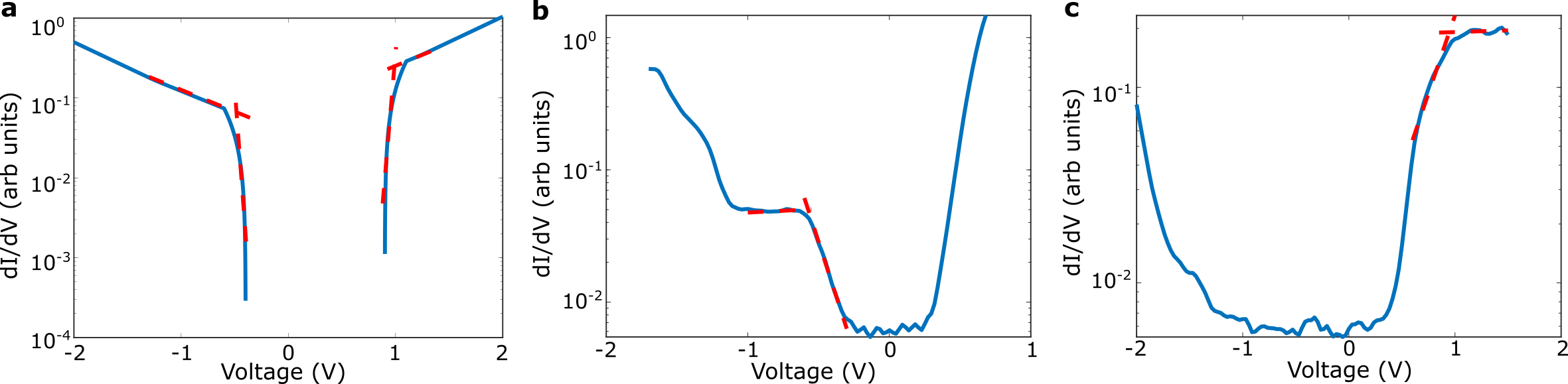}
\vspace{-0.15 in}
\caption{\small{ \textbf{a} Theoretical calculation of the temperature-broadened differential conductance of a semiconductor with valence band edge at -0.5 V and conduction band edge at +1.0 V. The dashed lines are linear fits to the spectrum above and below the band edges, and the crossing points mark the position of the band edges. \textbf{b,c} Determination of the valence and conduction band edges from typical experimental spectra following the procedure outlined in a.
}}
\captionsetup{labelformat=empty}
\vspace{-0.15 in}
\end{figure}

\begin{equation}
    \frac{dI(V)}{dV}=\int_{-\infty}^{+\infty} \! \rho_S(E+eV)\frac{df(E)}{dE} \, \mathrm{d}E
\end{equation}
where $f(E)$ is the Fermi-Dirac distribution at the temperature of the measurement. The result of this process is shown in figure S3a for a hypothetical semiconductor with a valence band edge at -0.5 eV and conduction band edge at 1.0 eV. The sharp band edge develops a finite slope due to temperature broadening at non-zero temperature. A simple practical method that we adopt to define the band edges is also shown on this plot, by drawing intersecting straight lines below and above the gap edge. The use of this method is shown in practice on real data in figures S3b and S3c for the valence and conduction bands respectively. We note that the accuracy of this process is not limited by the temperature broadening of the spectrum - the determination of the band edge is ultimately set by the signal to noise ratio of the measurement.

\section{S4. Analysis leading to figures 3d and 4a of the main text}

In order to extract band edges as a function of \moire wavelength, we start by defining the \moire unit cells, based upon the positions of the XX' positions from the topograph in figure 3a of the main text (see figure S4a). We avoid \moire unit cells that have aspect ratios larger than 2. We then use Delaunay triangulation to define the \moire unit cells from the XX' positions shown in figure S4a. For each Delaunay triangle, we then find the centroid, which defines the position of MM' and MX' stacking respectively. We use the spectroscopic data of figure 3b and 3c to find the band edges at these points, and set this to the valence and conduction band edge values for each given triangle. The result of this procedure for the valence band edge is shown in figure S4b. We then take the area of the \moire unit cell and convert it to a \moire wavelength by assuming it to be an equilateral triangle in order to generate the plot in figure 3d of the main text. This necessarily introduces scatter into the data, since triangles with different aspect ratios have different spectroscopic properties. However, it allows us to represent the entire data set of band edges as a function of \moire wavelength. 

In order to then find the \moire potential for each \moire unit cell, we take the difference in band edges between each MX' region and the three neighboring MM' regions, as shown in figure S4c. This data is used to generate the plot in figure 4a of the main text.

\begin{figure}[hbp]
\centering
\includegraphics[width=\textwidth]{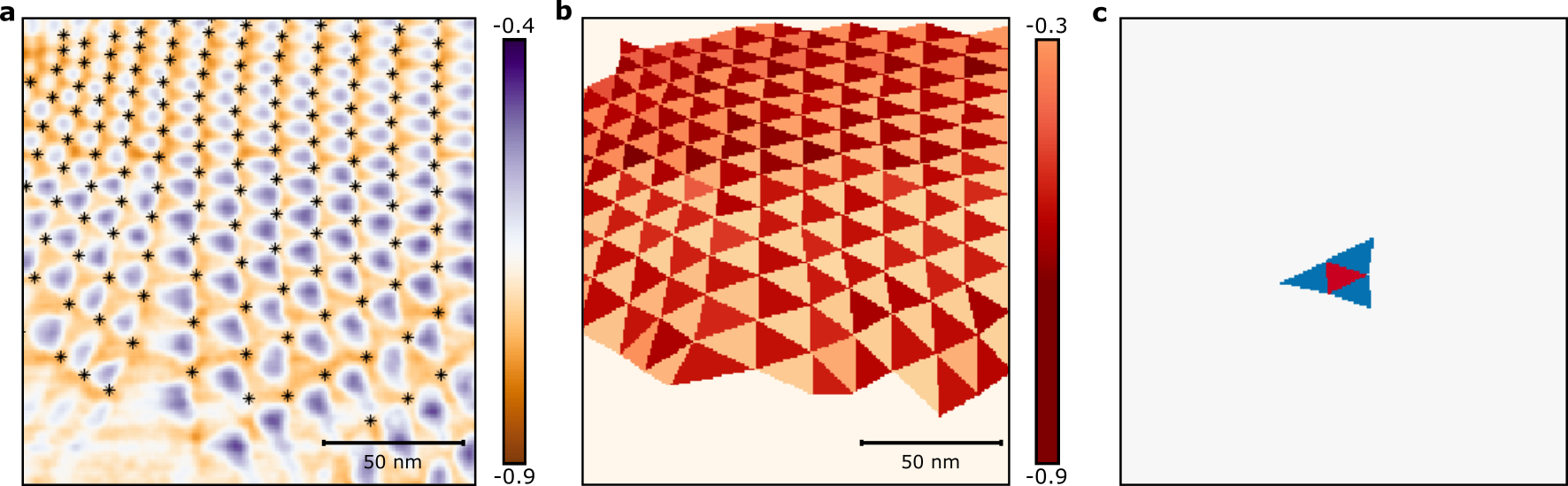}
\vspace{-0.15 in}
\caption{\small{ \textbf{a} XX' points of the \moire lattice shown as black markers, superposed on the spectroscopic map of valence band edges (in $eV$). \textbf{b} Determination of valence band edges of each MX' and MM' region after Delaunay triangulation of the XX' points in a (in $eV$). \textbf{c} Determination of the \moire potential. The \moire potential is defined as the absolute value of the difference between a given triangle in b (shown in red) and the average of its three nearest neighbors shown in blue.
}}
\captionsetup{labelformat=empty}
\vspace{-0.15 in}
\end{figure}